\documentclass[11pt]{article}
\usepackage[utf8]{inputenc}
\usepackage[T1]{fontenc}
\usepackage{cite}
\usepackage[a4paper,margin=1.2in]{geometry}
\usepackage{amsmath,amssymb}
\usepackage{graphicx}
\usepackage{xcolor}
\usepackage{hyperref}
\usepackage{xspace}
\hypersetup{colorlinks=true,linkcolor=blue,citecolor=magenta,urlcolor=blue}

\newcommand{\born}{\mathcal{B}}
\newcommand{\virtual}{\mathcal{V}}
\newcommand{\real}{\mathcal{R}}

\newcommand{\ppzz}{pp\rightarrow ZZ}

\newcommand{\mcatnlo}{MC@NLO\xspace}
\newcommand{\powheg}{POWHEG\xspace}
\newcommand{\krknlo}{KrkNLO\xspace}
\newcommand{\macnlops}{MAcNLOPS\xspace}
\newcommand{\esme}{ESME\xspace}

\title{\textbf{MAcNLOPS for $ZZ$ Pair Production at the LHC}}
\author{
Yuxiao Che\footnote{E-mail: \texttt{yu7648ch-s@student.lu.se}}
\; and
Rikkert Frederix\footnote{E-mail: \texttt{rikkert.frederix@fysik.lu.se}}
\;
\\
{\small\it Department of Physics, Lund University, Box 118, 221 00 Lund, Sweden}
}

\begin{document}

\date{}
\maketitle

\begin{abstract}
  \noindent We present an implementation of the \macnlops matching
  prescription for $\ppzz$ production in a
  \textsc{MadGraph5\_aMC@NLO}+\textsc{Pythia8} setup. Starting from a
  standard \mcatnlo event sample, negative $\mathbb{H}$ events are
  removed and compensated by a veto applied to the first shower
  emission of the $\mathbb{S}$ events. The implementation is validated
  against \mcatnlo for radiation-sensitive and inclusive diboson
  observables. Agreement is found up to a rather small
  power-suppressed contribution affecting the very low-$p_T$
  region. The method removes all negative $\mathbb{H}$ weights with
  negligible additional computational cost, while negative
  $\mathbb{S}$ weights are left unchanged, showing that \macnlops is a
  promising alternative to \mcatnlo with a reduced fraction of
  negative weights.
\end{abstract}

\section{Introduction}

Precision predictions for LHC observables are based on event
generators that combine fixed-order perturbative accuracy with fully
exclusive final states. In this context, NLO+PS matching is a baseline
requirement for phenomenological studies and experimental
analyses. The matching procedure must preserve the NLO accuracy of
inclusive observables, describe the first hard QCD emission reliably,
and at the same time remain compatible with the logarithmic structure,
recoil prescription and event structure of the parton shower. These
requirements are not independent, and different matching schemes
realise the compromise between fixed-order accuracy, shower control
and event-generation efficiency in different ways.

A canonical formulation of this problem is provided by the
\mcatnlo method \cite{Frixione:2002ik}. Its central idea is to
subtract from the real-emission matrix element the approximation
already generated by the shower, and to add the corresponding
integrated contribution to the Born-like event sample. This gives
NLO-accurate predictions while leaving the subsequent shower evolution
to the parton-shower program. The price paid for this generality is
the appearance of events with negative weights. Negative weights do
not invalidate the calculation: they are a consequence of the local
cancellation structure of the matched cross section. They are,
however, undesirable in practice. They reduce the statistical power of
event samples, increase the number of generated events needed for a
given precision, and complicate the use of simulated samples in
experimental analyses.

Several approaches have been proposed to address this problem. In the
\powheg method \cite{Nason:2004rx,Frixione:2007vw}, the hardest emission is
generated with a positive Sudakov formulation based on the real matrix
element. This largely avoids the negative-weight problem and has been
highly successful phenomenologically, but it also changes the division
of labour between the NLO calculation and the parton shower: the first
emission is generated by the matching framework rather than by the
shower itself. This distinction can be relevant when one wants to
preserve, as much as possible, the logarithmic structure and recoil
strategy of a given shower. Other methods, such as \krknlo
\cite{Jadach:2015mza}, also aim at positive-weight event samples, but
are less straightforward to apply as fully general matching
prescriptions.

Within the \mcatnlo framework itself, a detailed analysis of
the origin of negative weights led to the \mcatnlo-$\Delta$
prescription \cite{Frederix:2020trv}. This method modifies the
\mcatnlo matching formula by introducing a Sudakov-like factor
$\Delta$ in the $\mathbb{H}$ contribution, together with a
compensating term in the $\mathbb{S}$ contribution. The factor is
constructed from shower no-emission probabilities and suppresses
$\mathbb{H}$ events in regions where the parton shower gives an
efficient description of the radiation.  This reduces the negative
$\mathbb{H}$-event component while preserving the NLO expansion and
the hard-emission behaviour of the matched result. Consequently,
\mcatnlo-$\Delta$ can significantly improve the
event-generation efficiency, but it does not remove the
negative-weight problem altogether.

The \macnlops method \cite{Nason:2021xke} provides a different
route. It can be viewed as a minimal modification of the \mcatnlo
construction in which the regions where the shower approximation
exceeds the real matrix element are treated multiplicatively, while
the remaining regions retain an additive structure. In this way, the
negative $\mathbb{H}$-event contribution of \mcatnlo is removed and
compensated by a veto procedure applied after the first shower
emission. The method therefore keeps a key practical advantage of
\mcatnlo, namely that the first emission is still generated by the
shower, while addressing one of its main efficiency limitations. In
the implementation studied here, the focus is on this minimal
\macnlops mechanism and, in particular, on the removal of negative
$\mathbb{H}$ weights. Possible negative $\mathbb{S}$ weights are a
separate issue and are not eliminated by the procedure considered in
this work.

More recently, the \esme framework, ``Exponentiated Subtraction for
Matching Events'', was introduced \cite{vanBeekveld:2025lpz}. Like
\macnlops, it aims at positive-weight NLO+PS matching while retaining
a close connection to the parton shower, but it addresses the problem
more broadly by exponentiating the subtraction contribution associated
with the NLO normalisation and treating it through an event-by-event
accept--reject procedure. In this sense, \esme is conceptually more
comprehensive than the minimal \macnlops construction studied
here.

The aim of this paper is to present a first implementation of this
\macnlops prescription for $\ppzz$ production, using
\textsc{MadGraph5\_aMC@NLO} (\textsc{MG5\_aMC}) \cite{Alwall:2014hca}
interfaced to \textsc{Pythia8} \cite{Bierlich:2022pfr}, and to
validate it against standard \mcatnlo. The $\ppzz$ process is a useful
first application. Its Born-level final state is colour singlet, which
avoids the additional complications associated with final-state QCD
radiation from coloured particles, while being less trivial than a
$2\to 1$ colour-singlet process. It therefore offers a controlled but
meaningful environment in which to study the practical behaviour of
\macnlops.  We compare the two approaches for radiation-sensitive and
inclusive observables, both after the first shower emission and after
the subsequent shower evolution. Particular attention is paid to the
low-emission region, where the compensation of removed negative
$\mathbb{H}$ events is most delicate.

This paper is organised as follows. In Sects.~\ref{sec:macnlops} and
\ref{sec:implementation}, we review the formulation of \macnlops and
describe the implementation strategy used in this work. We present
numerical results and compare them to standard \mcatnlo in
Sect.~\ref{sec:results} and conclude in Sect.~\ref{sec:conclusions}.

\section{MAcNLOPS}
\label{sec:macnlops}

We first review the form of the \macnlops matching formula and then
describe how it is implemented in the present calculation. Throughout
this section, the formulae are written for a single real-emission
sector. Sums over flavours, FKS sectors and parton luminosities are
suppressed in the notation, but are included in the implementation. We
assume a factorisation of the real-emission phase space,
\begin{equation}
    d\Phi = d\Phi_B\,d\Phi_R ,
\end{equation}
where $\Phi_B$ denotes the underlying Born phase space and $\Phi_R$ the
radiation variables.

In \mcatnlo matching \cite{Frixione:2002ik}, the hard-event generation
can be written in terms of two event classes. The $\mathbb{S}$ events
have Born kinematics, while the $\mathbb{H}$ events have real-emission
kinematics, schematically,
\begin{align}
    \mathcal{F}^{(\mathbb{S})}(\Phi_B)
    &=
    \born(\Phi_B)
    + \virtual(\Phi_B)
    + \int \mathcal{K}(\Phi_B,\Phi_R)\,d\Phi_R
    \equiv
    \widetilde{\born}(\Phi_B),
    \label{eq:s_gen}
    \\
    \mathcal{F}^{(\mathbb{H})}(\Phi)
    &=
    \real(\Phi)-\mathcal{K}(\Phi).
    \label{eq:h_gen}
\end{align}
Here $\born$, $\virtual$ and $\real$ denote the Born, virtual and
real-emission contributions, respectively. The quantity $\mathcal{K}$
is the Monte Carlo counterterm, i.e., the shower approximation to the
real-emission matrix element in the corresponding singular region. It
removes the part of the real-emission contribution that would
otherwise be generated again by the parton shower.

The origin of negative $\mathbb{H}$ weights is apparent from
Eq.~\eqref{eq:h_gen}. Whenever the shower approximation locally
exceeds the real matrix element, $\mathcal{K}(\Phi)>\real(\Phi)$, the
$\mathbb{H}$-event weight is negative. This is not a pathology of the
matched cross section, since the negative contribution is compensated
by the showered $\mathbb{S}$ events, but it is inefficient for event
generation. The purpose of \macnlops is to explicitly perform this
cancellation between the $\mathbb{H}$ and $\mathbb{S}$ events at the
level of the hard cross section,
\begin{align}
    &\mathcal{F}^{(\mathbb{S}+\text{PS})}(\Phi) = \bigg(\widetilde{\born}(\Phi_B) \otimes \text{PS}_1(\Phi) \bigg) \otimes \mathcal{P}_\text{veto}(\Phi) \label{eq:s+ps} \\
    &\mathcal{F}^{(\mathbb{H})}_+(\Phi) = \big(\real(\Phi) - \mathcal{K}(\Phi)\big)\Theta(\real - \mathcal{K}), \label{eq:h}
\end{align}
by removing the negative contributions from the $\mathbb{H}$ events,
through the inclusion of the theta function in Eq.~\eqref{eq:h}, and
vetoing a corresponding fraction of the $\mathbb{S}$ events. In
Eq.\eqref{eq:s+ps}, for a Born configuration $\Phi_B$, the first
shower emission gives either no resolvable emission above the shower
cut-off or a real-emission phase-space point $\Phi$. Schematically,
the corresponding one-emission shower probability can be written as
\begin{equation}
    \text{PS}_1(\Phi)
    =
    \Delta(t_{\mathrm{cut}},\Phi_B)
    +
    \frac{\mathcal{K}(\Phi)}{\born(\Phi_B)}
    \Delta(t,\Phi_B)\,d\Phi_R ,
\end{equation}
where $\Delta(t,\Phi_B)$ is the shower Sudakov factor and $t$ is the
shower ordering variable. The first term corresponds to no emission
above the cut-off $t_{\mathrm{cut}}$, while the second term
corresponds to a first emission at scale $t$. The \macnlops
modification in Eq.\eqref{eq:s+ps} consists of accepting the one-step
showered $\mathbb{S}$ events with the probability
\begin{equation}
    \mathcal{P}_{\mathrm{veto}}(\Phi)
    =
    1+\frac{\real(\Phi)-\mathcal{K}(\Phi)}{\mathcal{K}(\Phi)}
    \Theta\!\left(\mathcal{K}(\Phi)-\real(\Phi)\right),
    \label{eq:veto}
\end{equation}
and removing the events from the sample otherwise.

Thus, in the region of the real-emission phase space where the shower
overestimates the real matrix element, the negative $\mathbb{H}$
events are explicitly removed by a theta function, and, at the same
time, a fraction of the $\mathbb{S}$ events is deleted after the first
shower emission, with a probability
$1-\mathcal{P}_{\mathrm{veto}}(\Phi)$. In the complementary region,
where the real matrix element is larger than the shower approximation
and event weights are already positive, the usual additive
$\mathbb{H}$ contribution is retained.  The resulting \macnlops cross
section is therefore
\begin{align}
    d\sigma_{\mathrm{MAcNLOPS}}
    &=
    \widetilde{\born}(\Phi_B)
    \Delta(t_{\mathrm{cut}},\Phi_B)\,d\Phi_B
    \nonumber \\
    &\quad
    +
    \widetilde{\born}(\Phi_B)
    \frac{\mathcal{K}(\Phi)}{\born(\Phi_B)}
    \Delta(t,\Phi_B)
    \left[
        1+
        \frac{\real(\Phi)-\mathcal{K}(\Phi)}{\mathcal{K}(\Phi)}
        \Theta\!\left(\mathcal{K}(\Phi)-\real(\Phi)\right)
    \right] d\Phi
    \nonumber \\
    &\quad
    +
    \left[
        \real(\Phi)-\mathcal{K}(\Phi)
    \right]
    \Theta\!\left(\real(\Phi)-\mathcal{K}(\Phi)\right)d\Phi .
    \label{eq:macnlops}
\end{align}
The last term is manifestly positive. For the second term, the factor
in square brackets is between zero and one, and does not introduce
(additional) negative $\mathbb{S}$ weights. Therefore, negative
$\mathbb{H}$ weights are removed. Possible negative $\mathbb{S}$
weights, originating from $\widetilde{\born}(\Phi_B)<0$, are not
addressed by this procedure, but can be reduced through
folding~\cite{Nason:2007vt,Frixione:2007nu,Alioli:2010xd,Frederix:2020trv}
and/or Born spreading~\cite{Frederix:2023hom}.

The NLO accuracy follows directly by comparing Eq.~\eqref{eq:macnlops}
with the standard \mcatnlo expression. In the region
$\real(\Phi)>\mathcal{K}(\Phi)$, the \macnlops expression is identical
to \mcatnlo. In the region $\mathcal{K}(\Phi)>\real(\Phi)$, their
difference is proportional to
\begin{equation}
    \left[
        \frac{\widetilde{\born}(\Phi_B)}{\born(\Phi_B)}
        \Delta(t,\Phi_B)-1
    \right]
    \left[
        \real(\Phi)-\mathcal{K}(\Phi)
    \right]
    \Theta\!\left(\mathcal{K}(\Phi)-\real(\Phi)\right).
\end{equation}
The difference $\real-\mathcal{K}$ is non-singular, since the two
terms have the same soft and collinear limits. The factor in the first
square bracket starts only beyond leading order. Hence the difference
between \macnlops and \mcatnlo is of NNLO size for infrared-safe
observables. The two prescriptions therefore have the same NLO
accuracy, while differing by matching-level higher-order terms.

\section{Implementation and validation}
\label{sec:implementation}

The implementation follows directly from
Eq.~\eqref{eq:macnlops}. Starting from a standard \mcatnlo event
sample generated with \textsc{MG5\_aMC} \cite{Alwall:2014hca} in the
default setup\footnote{\label{foot1}In the event generation we set
\texttt{UseSudakov = .false.} in \texttt{madfks\_mcatnlo.inc}. This
disables the built-in \textsc{MG5\_aMC} option that applies a
Sudakov-inspired damping to part of the $\mathbb{H}$ contribution, so
that the input sample corresponds to the standard additive \mcatnlo
structure of Eqs.~\eqref{eq:s_gen} and~\eqref{eq:h_gen}.}, the
negative $\mathbb{H}$ events are removed from the final sample and
their contribution is compensated by applying the veto probability in
Eq.~\eqref{eq:veto} to showered $\mathbb{S}$ events. The remaining
positive $\mathbb{H}$ events are kept unchanged. The showering is
performed with \textsc{Pythia8} \cite{Bierlich:2022pfr}.

For each $\mathbb{S}$-event, \textsc{Pythia8} is first run only up to
the first QCD shower emission. Since the process considered here has a
colour-singlet Born final state, only initial-state QCD radiation can
contribute at this stage. We therefore use a \textsc{Pythia8} user
hook that vetoes any subsequent shower emission after the first
accepted initial-state emission. Hadronisation, multiparton
interactions, QED showering and final-state radiation are switched off
during this intermediate step. Events with no resolvable emission
above the shower cut-off are kept without applying any veto.

For events with one shower emission, the real-emission configuration
is used to evaluate the ratio entering Eq.~\eqref{eq:veto}. This
requires the real matrix element $\real$ and the Monte Carlo
counterterm $\mathcal{K}$ at the emitted phase-space point. These
quantities can be evaluated with existing \textsc{MG5\_aMC} routines,
since the same quantities are also needed for the existing \mcatnlo
matching.

The veto probability is then computed according to
Eq.~\eqref{eq:veto}, i.e., events in the region $\mathcal{K}>\real$
are accepted with probability $\real/\mathcal{K}$, while events in the
region $\real\geq\mathcal{K}$ are always accepted. The accepted
$\mathbb{S}$ events are then combined with the positive $\mathbb{H}$
events to form the final \macnlops hard-event sample. This sample is
passed back to \textsc{Pythia8} for the remaining shower
evolution. For accepted $\mathbb{S}$ events with a first emission, the
subsequent shower is started from the scale of that emission.

\begin{figure}[t!]
    \centering
    \includegraphics[width=0.6\linewidth]{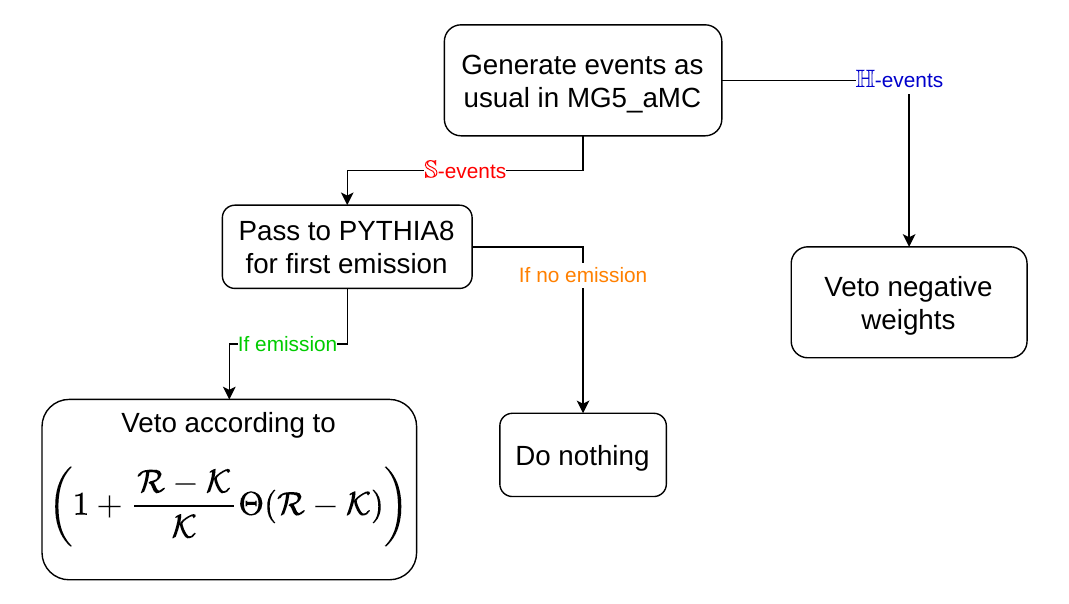}
    \caption{Schematic event-generation workflow used for the
      \macnlops implementation.}
    \label{fig:flowchart}
\end{figure}

The full event-generation chain can therefore be summarised as
follows:
\begin{enumerate}
    \item generate an \mcatnlo event sample with \textsc{MG5\_aMC};
    \item split the sample into $\mathbb{S}$ and $\mathbb{H}$ events;
    \item keep the positive $\mathbb{H}$ events and discard the
      negative $\mathbb{H}$ events;
    \item shower the $\mathbb{S}$ events up to the first QCD emission;
    \item apply the \macnlops veto of Eq.~\eqref{eq:veto} to the
      emitted $\mathbb{S}$ events;
    \item combine the accepted $\mathbb{S}$ events with the retained
      positive $\mathbb{H}$ events;
    \item run the remaining parton shower on the combined event
      sample.
\end{enumerate}
A schematic overview of the procedure is shown in
Fig.~\ref{fig:flowchart}.

\begin{figure}[t!]
    \centering
    \includegraphics[width=0.49\textwidth]{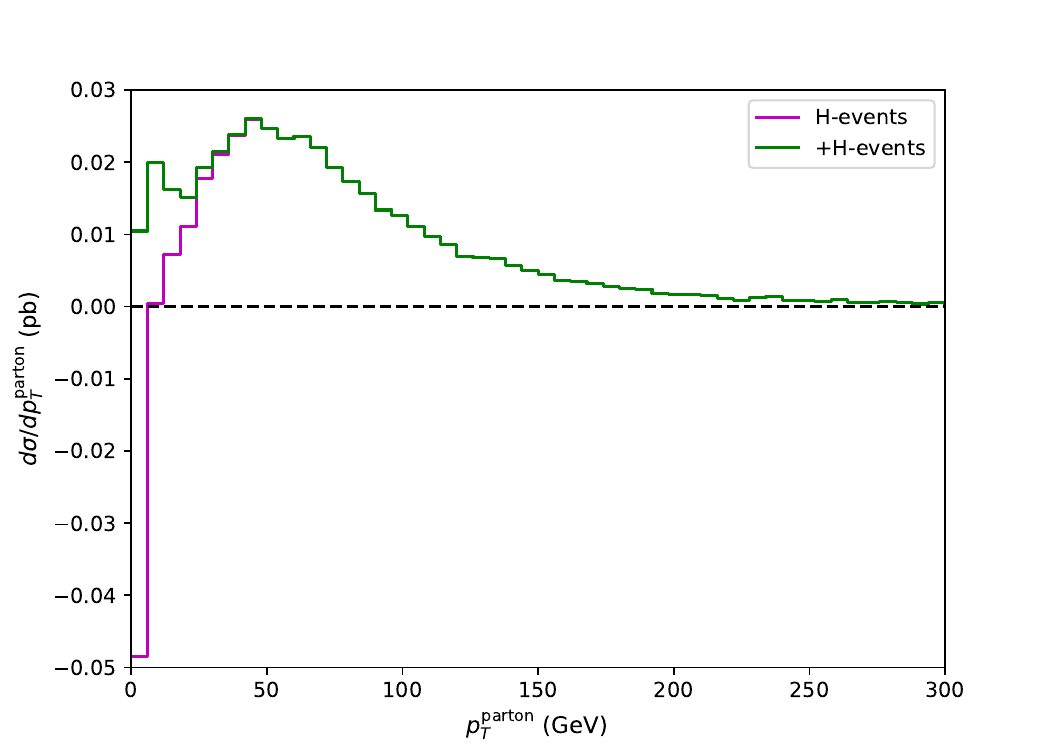}
    \includegraphics[width=0.49\textwidth]{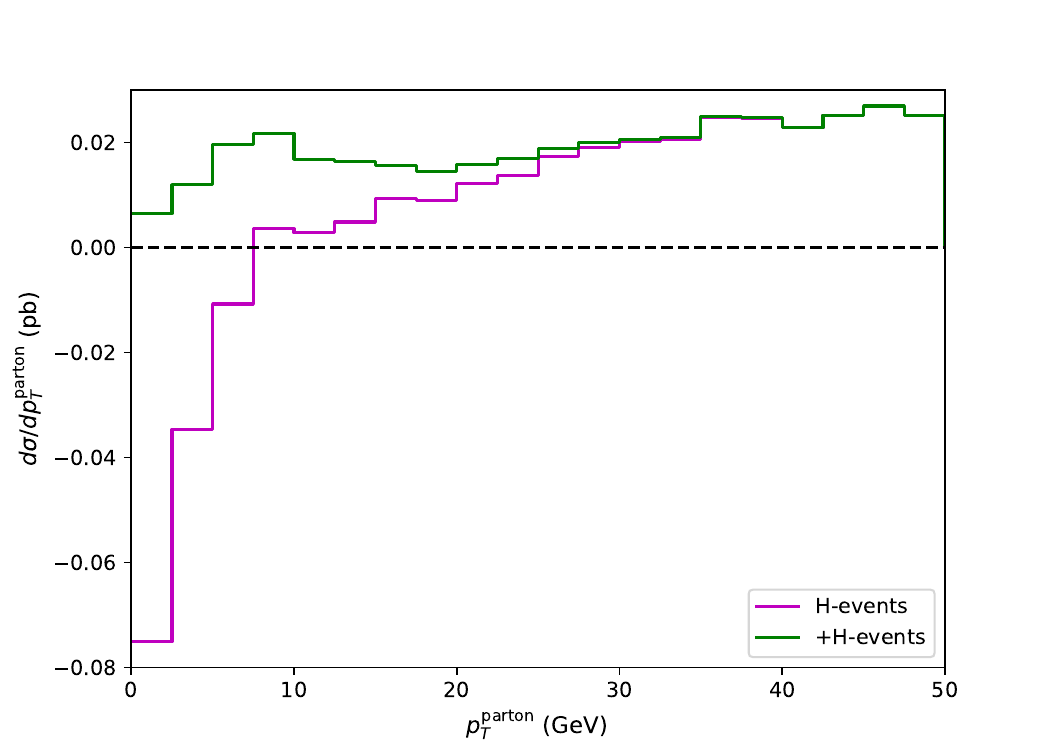}
    \caption{The distributions for the transverse momentum of the
      final-state parton in the $\mathbb{H}$ events without discarding
      the negative ones (magenta), and only positive $\mathbb{H}$
      events (green).}
    \label{fig:H_events_only}
\end{figure}

In order to validate the implementation of the \macnlops procedure for
$\ppzz$, we identify the phase-space regions where the negative
$\mathbb{H}$-event contributions arise.
Figure~\ref{fig:H_events_only} compares the full input
$\mathbb{H}$-event contribution (in magenta) with the contribution
obtained after retaining only positive $\mathbb{H}$ events (in green),
as a function of the transverse momentum of the final state parton
$p_T^{\mathrm{parton}}$ present in the $\mathbb{H}$ events before
showering. We remind the reader that, due to (transverse) momentum
conservation, at this level of predictions we have
$p_T^{\mathrm{parton}}=p_T^{ZZ}$.  In other words, the magenta curves
show the $\mathbb{H}$ events as present in \mcatnlo, and in green as
present in \macnlops.  The negative $\mathbb{H}$ events are
concentrated at small $p_T^{\mathrm{parton}}$, as expected from the
fact that they arise where the shower approximation can locally exceed
the real-emission matrix element. They are negligible in the hard
tail, become visible below roughly $p_T^{\mathrm{parton}}\simeq
30~\mathrm{GeV}$, and dominate the $\mathbb{H}$ contribution in the
lowest bins.

\begin{figure}[t!]
    \centering
    \includegraphics[width=0.67\linewidth]{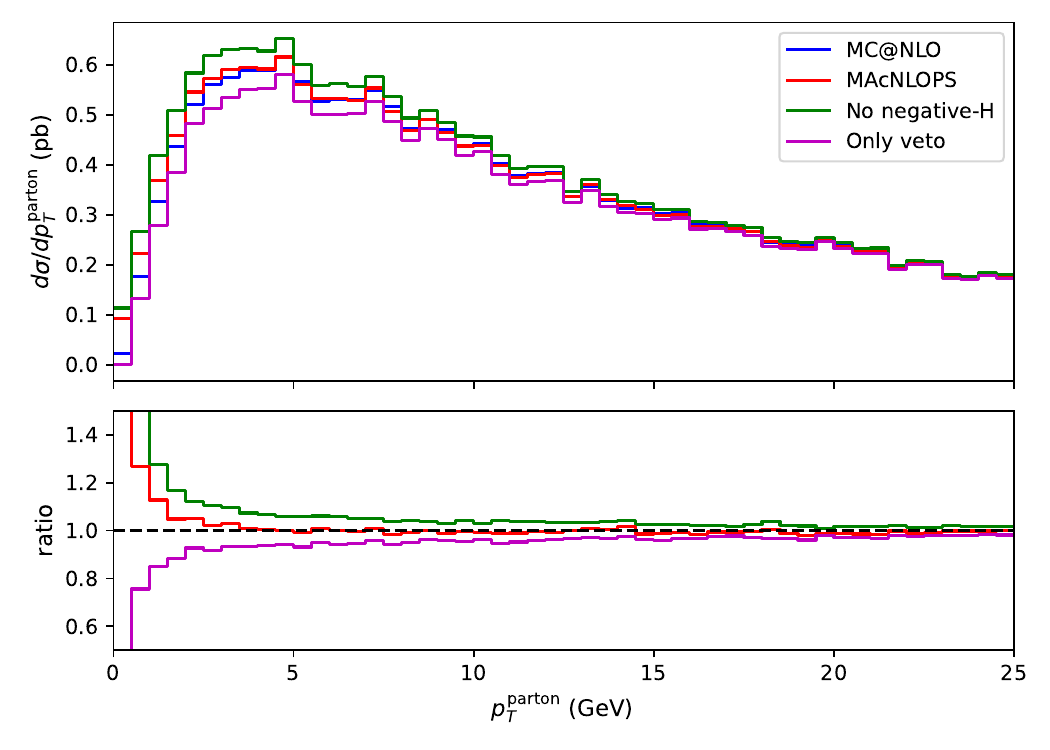}
    \caption{Histograms showing the separate effects of removing negative
    $\mathbb{H}$ events and \macnlops vetoing, with
    \macnlops and its input \mcatnlo for reference.}
    \label{fig:veto_vs_no_neg}
\end{figure}

This difference needs to be compensated for by the veto algorithm for
the $\mathbb{S}$ events, after a single emission by the shower. The
effect is isolated in Fig.~\ref{fig:veto_vs_no_neg} for the
low-$p_T^{\mathrm{parton}}$ range. The figure compares the sum of the
$\mathbb{H}$ events and the one-step-showered $\mathbb{S}$
events. Shown are the distribution for $p_T^{\mathrm{parton}}$ (i)
before removing negative $\mathbb{H}$ events and vetoing $\mathbb{S}$
events (this is equal to \mcatnlo and labeled as such in the plot);
(ii) only removing negative $\mathbb{H}$ events without applying the
veto (in green); (iii) only vetoing the $\mathbb{S}$ events (in
magenta); and (iv) the final \macnlops prediction (in red). In the
region where $\mathcal{K}>\mathcal{R}$, removing the negative
$\mathbb{H}$ term increases the distribution, while the veto on
showered $\mathbb{S}$ events decreases it. Away from the first few
bins, $p_T^{\mathrm{parton}}\gtrsim 2.5~\mathrm{GeV}$, these two
effects compensate each other at or below the 1\% level.

The behaviour in the first bins deserves separate comment. The
compensation for the removed negative $\mathbb{H}$ contribution is not
expected to be point-by-point exact. Expanding
Eq.~\eqref{eq:macnlops}, the difference between the \macnlops and
\mcatnlo expressions in the region $\mathcal{K}>\mathcal{R}$ is
proportional to
\begin{equation}
    \left[
        \frac{\widetilde{\born}}{\born}\Delta - 1
    \right]
    \left(\real-\mathcal{K}\right)
    \Theta(\mathcal{K}-\real) .
\end{equation}
The replacement $(\widetilde{\born}/\born)\Delta\simeq 1$ is
sufficient for NLO accuracy because $\real-\mathcal{K}$ is
non-singular, but it need not be numerically a small effect in the
immediate vicinity of the shower cut-off. In particular, negative
$\mathbb{H}$ events at an emission scale below the shower cut-off are
removed from the sample through the theta function in
Eq.~\eqref{eq:h}, but the corresponding
$\mathcal{P}_{\mathrm{veto}}(\Phi)$ for the $\mathbb{S}$ events is
only applied to events that have a shower emission, and are therefore
by construction above the shower cut-off. This mismatch is a
power-suppressed effect and can formally be neglected; it leads to the
\macnlops predictions having a slightly larger rate for
$p_T^{\mathrm{parton}}\gtrsim 2.5~\mathrm{GeV}$ than the \mcatnlo
reference.

We can conclude that the \macnlops implementation works as expected
for $\ppzz$ production. The compensation mechanism for the discarded
negative $\mathbb{H}$ events by the veto on the $\mathbb{S}$ events is
accurate at below the percent level, apart from the power-suppressed
region sensitive to the shower cut-off.

\section{Results}
\label{sec:results}

We now compare the \macnlops implementation for $\ppzz$ production at
the LHC to the default \mcatnlo prediction\footnote{Note that,
contrary to the previous section, for the \mcatnlo predictions, we use
the default \texttt{UseSudakov = .true.}  (see also footnote
\ref{foot1}), and therefore these events are no longer identical to
the input events to the \macnlops procedure.\label{foot2}}. The event
samples contain $10^5$ hard events at $\sqrt{s}=13~\mathrm{TeV}$,
generated in the five-flavour scheme with fixed renormalisation and
factorisation scales $\mu_R=\mu_F=\sqrt{\hat{s}_B}$, and parton
distribution functions are taken from the \textsc{CT18NLO} set
\cite{Hou:2019efy}. The subsequent showering is performed with the
default $p_T$-ordered shower of \textsc{Pythia8} with QED showering,
hadronisation and $Z$ decays switched off. The results are
therefore parton-level tests of the matching procedure.

The purpose of the comparison is to show that the \macnlops prediction
reproduces the corresponding \mcatnlo result within higher-order
matching-level effects. Unless explicitly stated otherwise, all
figures in this section have a similar layout: a main panel with the
reference \mcatnlo predictions in blue and the new \macnlops
predictions in red, while the lower panel contains the bin-by-bin
ratio of \macnlops with \mcatnlo. We do not include uncertainty
estimates and instead focus on the difference between the central
values of the predictions using equivalent input parameters.

\begin{figure}[t!]
    \centering
    \includegraphics[width=0.49\textwidth]{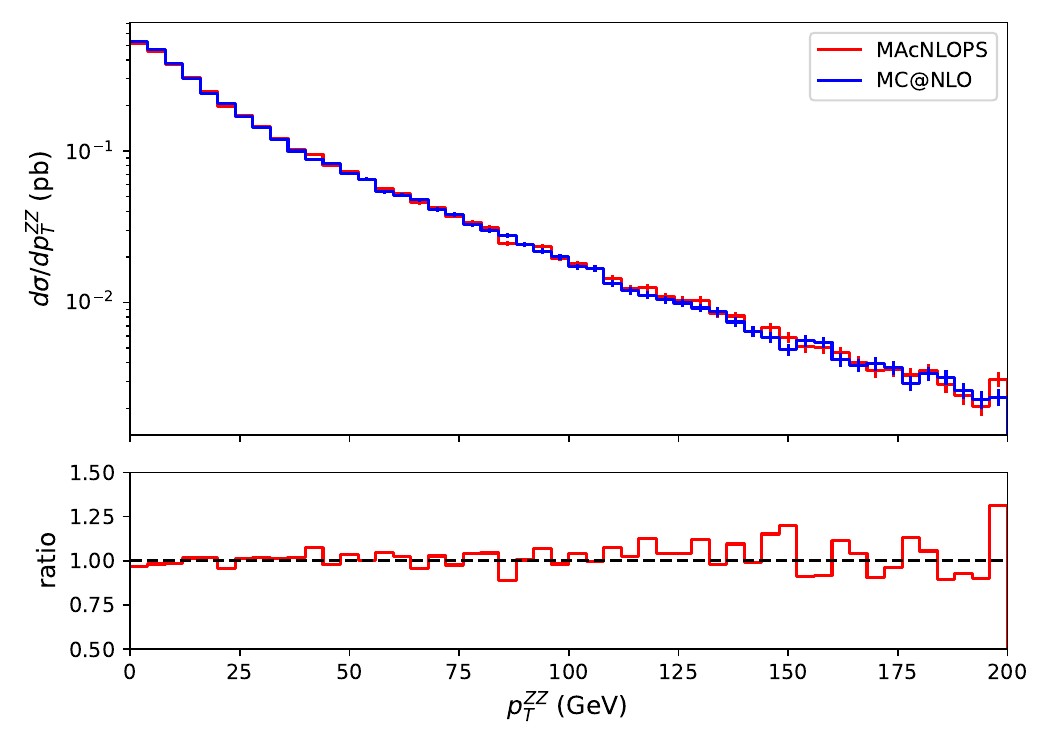}
    \includegraphics[width=0.49\textwidth]{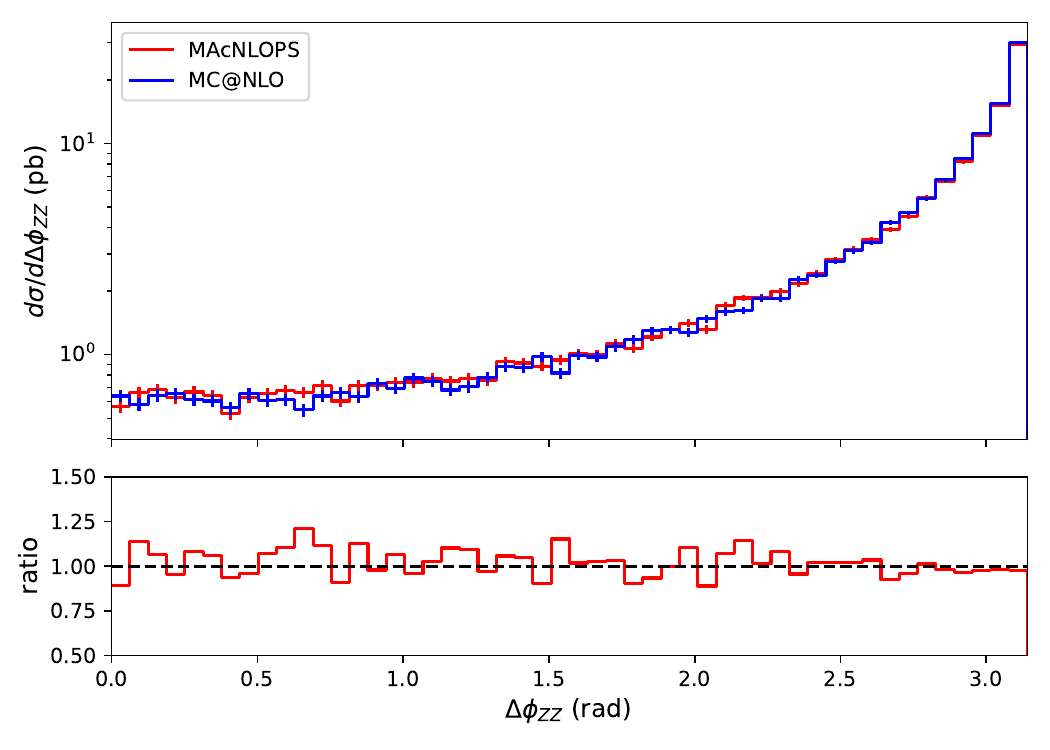}
    \caption{Histograms for the distribution of the transverse
      momentum of the $ZZ$ pair (left) and difference in azimuthal
      angle between the two $Z$ bosons (right) for \macnlops (in red),
      with \mcatnlo (in blue) as reference, showered with
      \textsc{Pythia8}.}
    \label{fig:azi_ZZ}
\end{figure}

The transverse momentum of the $ZZ$ pair is shown in
Fig.~\ref{fig:azi_ZZ} (left). The predictions for \macnlops and
\mcatnlo lie, essentially, on top of each other. The sensitivity to
the shower cut-off, as commented on in the discussion below
Fig.~\ref{fig:veto_vs_no_neg}, seems to have disappeared when showering
the events. This is not the full story, though. Rather, the \mcatnlo
prediction shown here, includes the \textsc{MG5\_aMC} option that
applies a Sudakov-inspired damping to the low-$p_T$ $\mathbb{H}$ event
contribution, see also footnotes~\ref{foot1} and~\ref{foot2}. This
removes some of the negative $\mathbb{H}$ events in the low $p_T$
region, increasing the \mcatnlo predictions in the first (couple of)
bins, and thereby bringing it closer to \macnlops.

A related radiation-sensitive observable is the azimuthal separation
of the two $Z$ bosons, shown in Fig.~\ref{fig:azi_ZZ} (right). At Born
level the two bosons are back-to-back, so the limit
$\Delta\phi_{ZZ}\to\pi$ probes the same recoil region as $p_T^{ZZ}\to
0$. The \macnlops and \mcatnlo predictions agree within statistical
uncertainties.

\begin{figure}[t!]
    \centering
    \includegraphics[width=0.49\textwidth]{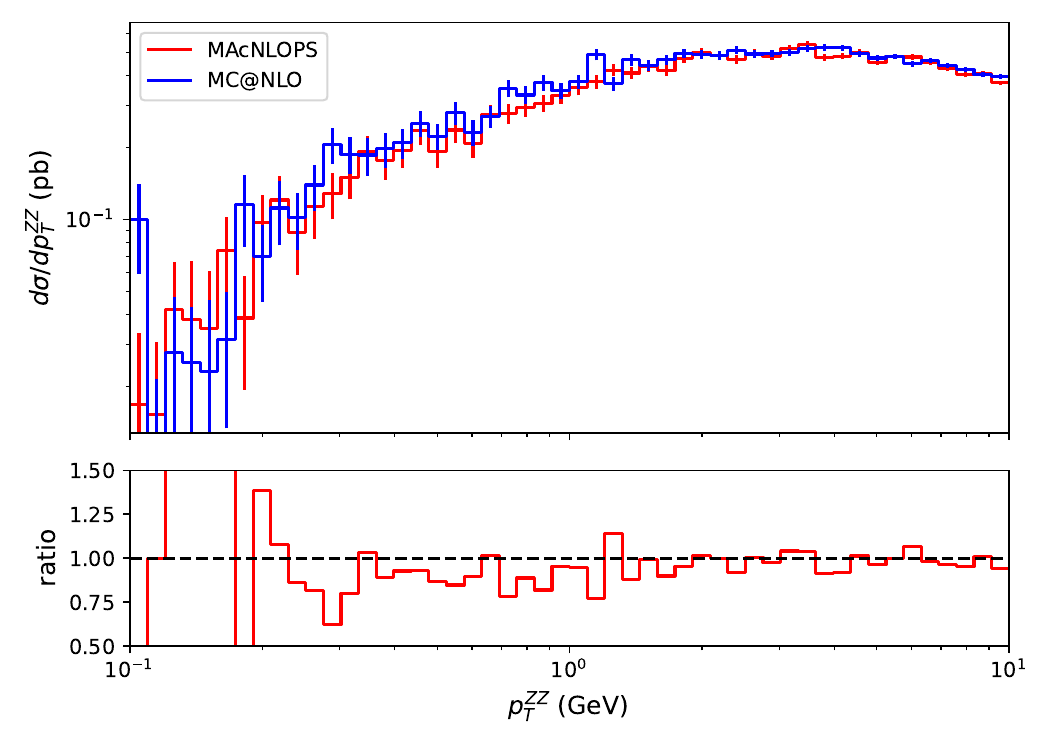}
    \includegraphics[width=0.49\textwidth]{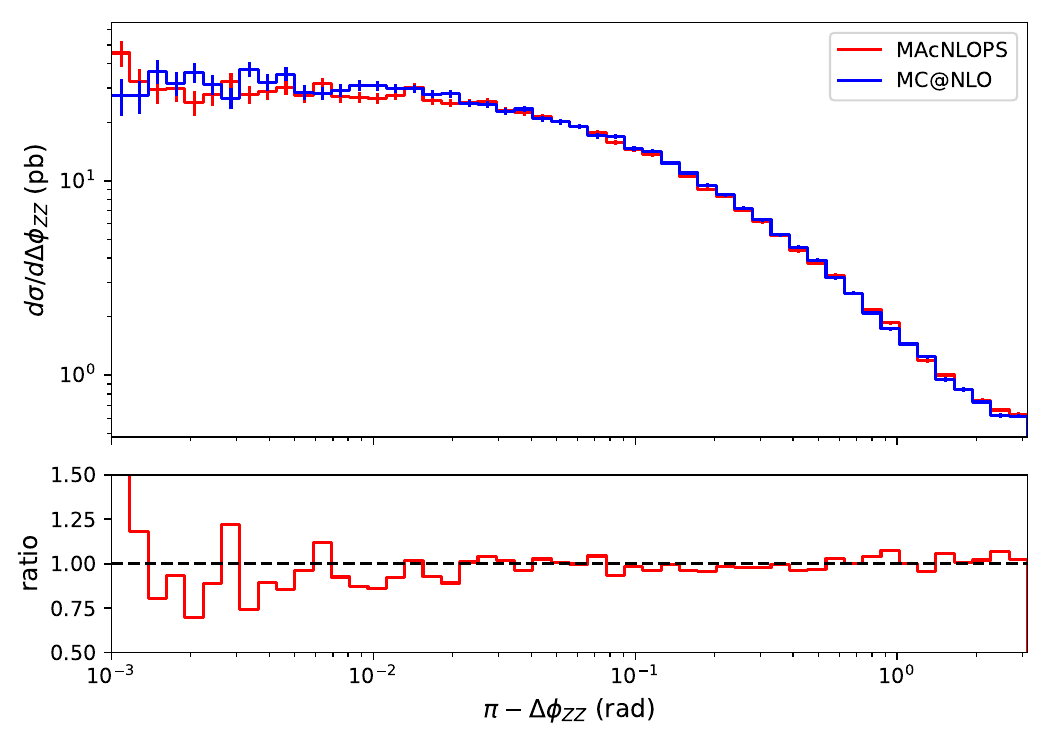}
    \caption{Histograms for $Z$ boson pair transverse momentum (left)
      and difference in azimuthal angle of the $ZZ$-pair (right) for
      \macnlops, with \mcatnlo and showered with
      \textsc{Pythia8}. Logarithmic binning is employed to emphasize
      the soft region.}
    \label{fig:loglog}
\end{figure}

In Fig.~\ref{fig:loglog} we show the $ZZ$ transverse momentum and the
azimuthal separation of the $ZZ$ system, using logarithmic binning to
emphasise the low-scale region. For the azimuthal observable we plot
the distribution in $\pi-\Delta\phi_{ZZ}$, so that the back-to-back
limit is displayed logarithmically. The \macnlops result follows the
\mcatnlo prediction within statistical uncertainties. A small deficit
of \macnlops relative to \mcatnlo is visible in the very soft region,
related to a power-suppressed dependence on shower cut-off and the
Sudakov-inspired damping at low-$p_T$ in the \textsc{MG5\_aMC}
implementation of \mcatnlo, as explained above.

\begin{figure}[t!]
    \centering
    \includegraphics[width = 0.49\textwidth]{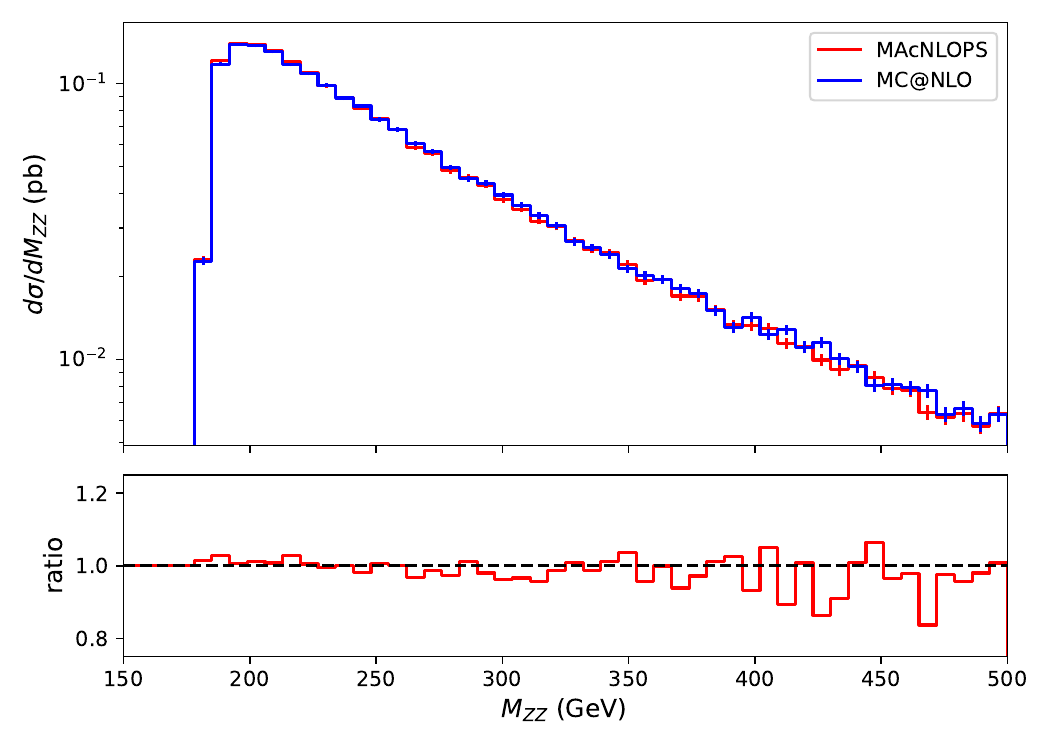}
    \includegraphics[width = 0.49\textwidth]{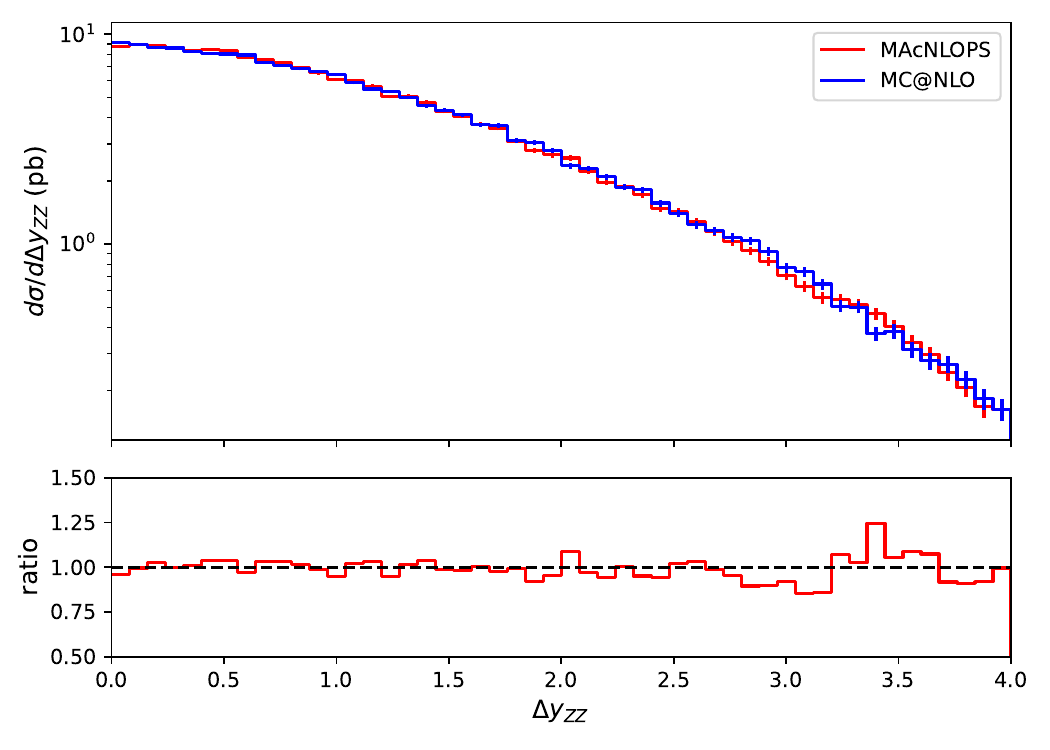}
    \caption{Histograms for invariant mass (left) and rapidity
      separation (right) of the $ZZ$-pair obtained with \macnlops
      (red), with \mcatnlo (blue) as reference.}
    \label{fig:mac_inclusive}
\end{figure}

Finally, the matching procedure should also leave Born-dominated
observables unchanged up to higher-order matching
effects. Figure~\ref{fig:mac_inclusive} shows the invariant mass and
rapidity separation of the $ZZ$ system. These observables are
primarily controlled by the underlying Born kinematics and are
therefore useful checks that the veto does not bias the hard
process. The \macnlops and \mcatnlo predictions agree within
statistical uncertainties for both observables. This provides a
consistency check of the event handling: the additional first-emission
processing does not produce a visible distortion of inclusive diboson
kinematics.

\section{Conclusions}
\label{sec:conclusions}

We have presented a first implementation of the \macnlops matching
prescription for $\ppzz$ production in a
\textsc{MG5\_aMC}+\textsc{Pythia8} setup. The implementation follows
the minimal construction proposed in Ref.~\cite{Nason:2021xke}:
negative $\mathbb{H}$ events are removed from the hard-event sample,
while their contribution is compensated by applying a veto to the
$\mathbb{S}$ events after the first shower emission. The subsequent
shower evolution is then performed with \textsc{Pythia8}.

The implementation was validated by comparing \macnlops to standard
\mcatnlo for several observables in $ZZ$
production. Radiation-sensitive distributions, such as the $ZZ$
transverse momentum, and the azimuthal separation of the two $Z$
bosons, show agreement between the present event samples. Inclusive
diboson observables, such as the invariant mass and rapidity
separation of the two $Z$ bosons, are also unchanged. These
comparisons provide a non-trivial check that the veto procedure
compensates the removed negative $\mathbb{H}$ contribution without
producing visible distortions in the matched prediction.

Special attention was paid to the low-scale region, where the
cancellation between the removed negative $\mathbb{H}$ contribution
and the vetoed $\mathbb{S}$ contribution is most delicate. The
\macnlops result remains compatible with the \mcatnlo prediction up to
power-suppressed effects. This indicates that the \macnlops
veto does not introduce an uncontrolled parton-level shape distortion
close to the shower cut-off.

The main practical outcome of this study is that the negative
$\mathbb{H}$ weights are removed without introducing a numerically
significant extra cost in the event-generation chain. The method does
not address negative $\mathbb{S}$ weights, which can still occur when
$\widetilde{\born}(\Phi_B)$ is negative. Combining \macnlops with
existing methods for reducing negative $\mathbb{S}$ weights, such as
folding~\cite{Nason:2007vt,Frixione:2007nu,Alioli:2010xd,Frederix:2020trv}
and/or Born spreading~\cite{Frederix:2023hom}, is therefore a natural
direction for future work.

The present implementation is tailored to a colour-singlet final
state, where the first QCD emission is purely initial-state
radiation. It should therefore be directly applicable, or require only
minor modifications, for similar colour-singlet production processes
with no coloured particles at Born level.  Extending the method to
processes with final-state radiation or more complicated colour
structures will require a more careful treatment of the shower
history, the phase-space mapping and the evaluation of the
corresponding Monte Carlo counterterms. Such extensions would be
important steps towards assessing whether \macnlops can provide a
general alternative to standard \mcatnlo with a substantially reduced
fraction of negative weights.

\bibliographystyle{JHEP}
\bibliography{main}

\providecommand{\href}[2]{#2}\begingroup\raggedright\begin{thebibliography}{10}

\bibitem{Frixione:2002ik}
S.~Frixione and B.R.~Webber, \emph{{Matching NLO QCD computations and parton
  shower simulations}},
  \href{https://doi.org/10.1088/1126-6708/2002/06/029}{\emph{JHEP} {\bfseries
  06} (2002) 029} [\href{https://arxiv.org/abs/hep-ph/0204244}{{\ttfamily
  hep-ph/0204244}}].

\bibitem{Nason:2004rx}
P.~Nason, \emph{{A New method for combining NLO QCD with shower Monte Carlo
  algorithms}},
  \href{https://doi.org/10.1088/1126-6708/2004/11/040}{\emph{JHEP} {\bfseries
  11} (2004) 040} [\href{https://arxiv.org/abs/hep-ph/0409146}{{\ttfamily
  hep-ph/0409146}}].

\bibitem{Frixione:2007vw}
S.~Frixione, P.~Nason and C.~Oleari, \emph{{Matching NLO QCD computations with
  Parton Shower simulations: the POWHEG method}},
  \href{https://doi.org/10.1088/1126-6708/2007/11/070}{\emph{JHEP} {\bfseries
  11} (2007) 070} [\href{https://arxiv.org/abs/0709.2092}{{\ttfamily
  0709.2092}}].

\bibitem{Jadach:2015mza}
S.~Jadach, W.~P{\l}aczek, S.~Sapeta, A.~Si{\'o}dmok and M.~Skrzypek,
  \emph{{Matching NLO QCD with parton shower in Monte Carlo scheme
  {\textemdash} the KrkNLO method}},
  \href{https://doi.org/10.1007/JHEP10(2015)052}{\emph{JHEP} {\bfseries 10}
  (2015) 052} [\href{https://arxiv.org/abs/1503.06849}{{\ttfamily
  1503.06849}}].

\bibitem{Frederix:2020trv}
R.~Frederix, S.~Frixione, S.~Prestel and P.~Torrielli, \emph{{On the reduction
  of negative weights in MC@NLO-type matching procedures}},
  \href{https://doi.org/10.1007/JHEP07(2020)238}{\emph{JHEP} {\bfseries 07}
  (2020) 238} [\href{https://arxiv.org/abs/2002.12716}{{\ttfamily
  2002.12716}}].

\bibitem{Nason:2021xke}
P.~Nason and G.P.~Salam, \emph{{Multiplicative-accumulative matching of NLO
  calculations with parton showers}},
  \href{https://doi.org/10.1007/JHEP01(2022)067}{\emph{JHEP} {\bfseries 01}
  (2022) 067} [\href{https://arxiv.org/abs/2111.03553}{{\ttfamily
  2111.03553}}].

\bibitem{vanBeekveld:2025lpz}
M.~van Beekveld, S.~Ferrario~Ravasio, J.~Helliwell, A.~Karlberg, G.P.~Salam,
  L.~Scyboz et~al., \emph{{Logarithmically-accurate and positive-definite NLO
  shower matching}}, \href{https://doi.org/10.1007/JHEP10(2025)038}{\emph{JHEP}
  {\bfseries 10} (2025) 038}
  [\href{https://arxiv.org/abs/2504.05377}{{\ttfamily 2504.05377}}].

\bibitem{Alwall:2014hca}
J.~Alwall, R.~Frederix, S.~Frixione, V.~Hirschi, F.~Maltoni, O.~Mattelaer
  et~al., \emph{{The automated computation of tree-level and next-to-leading
  order differential cross sections, and their matching to parton shower
  simulations}}, \href{https://doi.org/10.1007/JHEP07(2014)079}{\emph{JHEP}
  {\bfseries 07} (2014) 079} [\href{https://arxiv.org/abs/1405.0301}{{\ttfamily
  1405.0301}}].

\bibitem{Bierlich:2022pfr}
C.~Bierlich et~al., \emph{{A comprehensive guide to the physics and usage of
  PYTHIA 8.3}},
  \href{https://doi.org/10.21468/SciPostPhysCodeb.8}{\emph{SciPost Phys.
  Codeb.} {\bfseries 2022} (2022) 8}
  [\href{https://arxiv.org/abs/2203.11601}{{\ttfamily 2203.11601}}].

\bibitem{Nason:2007vt}
P.~Nason, \emph{{MINT: A Computer program for adaptive Monte Carlo integration
  and generation of unweighted distributions}},
  \href{https://arxiv.org/abs/0709.2085}{{\ttfamily 0709.2085}}.

\bibitem{Frixione:2007nu}
S.~Frixione, P.~Nason and G.~Ridolfi, \emph{{The POWHEG-hvq manual version
  1.0}},  \href{https://arxiv.org/abs/0707.3081}{{\ttfamily 0707.3081}}.

\bibitem{Alioli:2010xd}
S.~Alioli, P.~Nason, C.~Oleari and E.~Re, \emph{{A general framework for
  implementing NLO calculations in shower Monte Carlo programs: the POWHEG
  BOX}}, \href{https://doi.org/10.1007/JHEP06(2010)043}{\emph{JHEP} {\bfseries
  06} (2010) 043} [\href{https://arxiv.org/abs/1002.2581}{{\ttfamily
  1002.2581}}].

\bibitem{Frederix:2023hom}
R.~Frederix and P.~Torrielli, \emph{{A new way of reducing negative weights in
  MC@NLO}}, \href{https://doi.org/10.1140/epjc/s10052-023-12243-x}{\emph{Eur.
  Phys. J. C} {\bfseries 83} (2023) 1051}
  [\href{https://arxiv.org/abs/2310.04160}{{\ttfamily 2310.04160}}].

\bibitem{Hou:2019efy}
T.-J.~Hou et~al., \emph{{New CTEQ global analysis of quantum chromodynamics
  with high-precision data from the LHC}},
  \href{https://doi.org/10.1103/PhysRevD.103.014013}{\emph{Phys. Rev. D}
  {\bfseries 103} (2021) 014013}
  [\href{https://arxiv.org/abs/1912.10053}{{\ttfamily 1912.10053}}].

\end{thebibliography}\endgroup
\end{document}